\documentclass[aps,preprint,superscriptaddress,showpacs]{revtex4}

\usepackage{graphicx,epsfig}

\def\be{\begin{equation}}
\def\ee{\end{equation}}
\def\bea{\begin{eqnarray}}
\def\eea{\end{eqnarray}}
\def\half{\frac{1}{2}}

\def\case#1/#2{\textstyle\frac{#1}{#2}}
\def\k0{\kappa_{0}}

\begin{document}

\title{Generating  Generalized $G_{D-2}$ solutions}
\author{N. Bret\'on$^1$, A. Feinstein$^2$ and L. A. L\'opez$^{}$}
\affiliation{$^{}$ Dpto. de F\'{\i}sica, Centro de Investigaci\'on y de
Estudios Avanzados del I. P. N.,
Apdo. 14-740, D.F., M\'exico.\\
$^2$ Dpto. de F\'{\i}sica Te\'orica, Universidad del Pa\'{\i}s
Vasco, Apdo. 644, E-48080, Bilbao, Spain.}

\begin{abstract}
We show how one can systematically construct vacuum solutions to Einstein
field equations with $D-2$ commuting Killing vectors in $D>4$ dimensions.
The construction uses Einstein-scalar field seed solutions in 4 dimensions
and is performed both for the case when all the Killing directions are
spacelike, as well as when one of the Killing vectors is timelike. The
later case corresponds to generalizations of stationary axially symmetric
solutions to higher dimensions. Some examples representing generalizations
of known higher dimensional stationary solutions are discussed in terms of
their rod structure and horizon locations and deformations.
\end{abstract}


\maketitle

\section{Introduction}
 
There has been a renewed interest in higher dimensional solutions to
Einstein field equations.  Several interesting vacuum static/stationary
solutions \cite{emparan}, the methods of their generations \cite{5dims}
and study \cite{harmark}, as well as some general results on uniqueness of
higher dimensional static vacuum black holes \cite{gibbons},\cite{wald},
\cite{hollands} in $5$ dimensions have appeared recently.  In the case of
time-dependent geometries, the main interest is to study higher
dimensional spacetimes as backgrounds for string propagation
\cite{aharony}. In cosmology \cite{cosmology}, the new trends impose
``lifting" the cosmological models to higher dimensions. Another, and
probably the most relevant reason to study the higher dimensional
generalizations of the Einstein equations stems from the fact that we live
in a four dimensional world. It would be important then, if we were able
to convince ourselves, by studying the higher dimensional solutions, that
there is something special, unique and deep about four dimensions. This
could only be done if we study the alternatives.

Much work has been done previously \cite{5d}, yet, since the interests
move with time, the motivation, and with it the boundary/initial
conditions imposed on the solutions change as well.  In solving Einstein
equations, one usually imposes some sort of symmetry. What we are good at
is the situation where the spacetime does not depend on more than two,
preferably non-null variables. In four dimensions these are known as the
so-called $G_{2}$ solutions. Almost all known interesting four-dimensional
solutions of Einstein equations in vacuum, electrovacuum or with some
fundamental matter fields belong to this class and contain at least two
commuting Killing directions. Once we are confronted with the situation
where the line element {\em does} depends on at most two coordinates, we
are in ``business": whether these are stationary solutions with axial
symmetry, boost-symmetric spacetimes, anisotropic (isotropic) and
inhomogeneous (homogeneous) cosmologies, cylindrical, or plane gravity and
matter waves - there are dozens of generating techniques, algorithms etc.
to construct new solutions of ever increasing complexity \cite{g2}
starting from more simple seeds. Because of their generality, on one hand,
and applicability, on the other, the known $G_{2}$ solutions provide a
perfect ``seed" or a building block to construct further new solutions.
The construction of families of higher dimensional solutions is not an
exception: one can efficiently use the known $G_{2}$ solutions as the
seeds in order to construct their higher dimensional generalizations and
analogues. As usual, the physics enters via the boundary and initial
conditions (asymptotic flatness, clean horizons and singularities in the
case of compact objects; singularity structure, inflation, late/early-time
acceleration-for cosmological models etc.) Obtaining sufficiently general
solutions, of course, does not necessarily mean alleviating the search for
specialty and physical meaningfulness, nevertheless, the understanding of
the sufficiently general class of solutions may shed some light on a
problem.

The above considerations lead us to the main purpose of this paper: the
construction, in a possibly most simple and controllable way of higher
dimensional vacuum solutions to Einstein equations which depend, at most,
on two variables.  Our starting point would be the $4$-dimensional vacuum
seed metrics with the $G_{2}$ symmetry .  The vacuum seed spacetime will
be generalized to include massless dilatons, which will serve to lift the
solutions to higher dimensions. To include the scalar field, the vacuum
line element would be written in the coordinates especially adopted for
this matter. It happens so, that in these coordinates the scalar dilaton
coupled to gravity in the $G_{2}$ case satisfies a linear differential
equation. This equation can be easily solved with the general solution
representing a linear combination of some elementary solutions. Any term,
as well as, any linear combination of these solutions, may serve to lift
the dilaton spacetime to higher dimensions, obtaining in such a way a new
vacuum solution in any dimension greater than $4$. To get interesting
solutions, one must choose the ``right" $G_{2}$ seed as well as the right
combination of elementary scalar solutions (dilatons). The scheme works
both, in the case when the two commuting Killing vectors of the ``seed"
are spacelike (cosmologies, cylindrical waves, colliding waves etc.) as
well as, when one of the Killing vectors is timelike. It is important to
mention that there is no need in imposing the hypersurface orthogonality
of the two Killing directions (diagonality of the metric) and therefore,
the solutions obtained this way generalize, in the case when one of the
Killing directions is timelike, for example, the static higher dimensional
solutions due to \cite{emparan}, enabling to obtain the generalized
stationary solutions, in fact, an infinite dimensional family of such
solutions.

In the following section we briefly review the generating technique for
the case where the two Killing directions of the seed solution are both
spacelike. In Sec. III the case with one timelike Killing vector is
addressed. In Sec. IV the method is specialized to the 5-D case; aspects
such as asymptotic flatness and the interpretation of the method as the
insertion of rod sources are as well included in this section. In Sec. V,
we analyze the trapping of surfaces, horizons and singularities of the
generated solutions. In Secs. VI and VII some 5-D static and stationary
solutions are generated. Finally, some conclusions are drawn in the last
section.

\section{All Spacelike Killing Vectors}

The case and the algorithm where the two Killing directions of the seed
solution are both spacelike, was first presented and discussed in the
context of string/M-theory cosmology \cite{feinmavm}. It is worth,
however, to briefly review it here since actually passing to the situation
when one of the Killing vectors is timelike, can be formally made by
complex coordinate changes which will be given subsequently.

The starting point is the {\em vacuum solution} of the Einstein field
equations in the form

\be
\label{er}
ds^2=e^{f^{\rm vac}}(-dt^2+dz^2)+\gamma_{ab}\,dx^a\,dx^b.
\ee

Here $f^{\rm vac}$ and $\gamma_{ab}$ are functions of $t$ and $z$
coordinates alone, $(x,y)\equiv(x^{2},x^{3})$ and we denote
$\sqrt{\det \gamma}\equiv K(t,z)$.

We also assume that the scalar field is normalized as in \cite{feinmavm}
and $\Phi\equiv \sum_{i=1}^{N}\varphi_{i}$ expressed as a sum of
elementary scalar fields solves the following linear differential equation
\be \label{KG1} \frac{\partial}{\partial t}[K(t,z)\dot{\Phi}(t,z)]-
\frac{\partial}{ \partial z}[K(t,z)\Phi'(t,z)] =0. \ee

Then, the solution to the coupled Einstein-scalar field equations is
obtained \cite{letelier} by keeping the transverse part characterized by
the metric functions $K(t,z)$ and $\gamma_{ab}$ without being changed, but
replacing the longitudinal function $f(t,z)^{\rm vac}$ by

\be f(t,z)^{\rm
vac} \longrightarrow f(t,z)^{\rm vac} + f(t,z)^{\rm sc}. \ee
The function $f(t,z)^{\rm sc}$, then, is solved by quadratures from:

\be
\dot{f}(t,z)^{\rm sc}=\frac{K}{{K'^{2}-\dot{K}^2}}\left[
2K'\sum_{i=1}^{N}\dot{\varphi}_{i}\varphi_{i}'
-\dot{K}\left(\sum_{i=1}^{N}\dot{\varphi}_{i}^{2}+
\sum_{i=1}^{N}\varphi_{i}'^{2}\right)\right],
\ee

\be
f'(t,z)^{\rm sc} = \frac{K}{{K'^{2}-\dot{K}^2}}\left[
K'\left(\sum_{i=1}^{N}\dot{\varphi}_{i}^{2}+
\sum_{i=1}^{N}\varphi_{i}'^{2}\right)
- 2\dot{K}\sum_{i=1}^{N}\dot{\varphi}_{i}\varphi_{i}'
\right].
\ee

To lift the solution to higher dimensions and to obtain the vacuum
spacetime, we first construct the new scalars \cite{feinmavm}
$\psi_i={\cal D}_{ij}\varphi_{j}$, where ${\cal D}_{ij}\in GL(N,{\bf R})$
is given by

\be
{\cal D}=\left(\matrix{\mu_{1}^{-\half} & \mu_{2}^{-\half} & \mu_{3}^{-\half}
& \ldots & \mu_{N-1}^{-\half} & \mu_{N}^{-\half} \cr
-\mu_{1}^{-\half} & \mu_{2}^{-\half} & \mu_{3}^{-\half} & \ldots &
\mu_{N-1}^{-\half} & \mu_{N}^{-\half} \cr
0 & -2\mu_{2}^{-\half} & \mu_{3}^{-\half} & \ldots &
\mu_{N-1}^{-\half} & \mu_{N}^{-\half} \cr
\vdots & \vdots & \vdots & & \vdots & \vdots \cr
0 & 0 & 0 & \ldots & \mu_{N-1}^{-\half} & \mu_{N}^{-\half} \cr
0 & 0 & 0 & \ldots & -(N-1)\mu_{N-1}^{-\half} & \mu_{N}^{-\half}}
\right),
\ee

along with

\begin{eqnarray}
\mu_{n}=&\frac{2}{3}n(n+1), \\
\mu_{N}=&\frac{1}{3}N(N+2),
\end{eqnarray}

where $n=1,\ldots,N-1$.

Finally, the $N$-dimensional vacuum solution is given by:

\be
\label{higher}
ds^2_{4+N} =e^{-\frac{2}{\sqrt{3}}\sum_{i=1}^{N}\psi_{i}}\, ds^{2}_{4}
+\sum_{i=1}^{N}e^{\frac{4}{\sqrt{3}}\psi_{i}} (dw^{i})^2,
\ee
where $ds^{2}_{4}$ is the four-dimensional scalar field solution
constructed previously. The new scalars $\psi$ need to be constructed only
when one is seeking a solutions in more than $5$ dimensions. In the 5
dimensional case the lifting from 4 dimensions involves directly the field
$\Phi$ and is straightforward.

The case of all spacelike Killing vectors is phenomenologically rich.
Depending on the behavior of the function $K(t,z)$, the so-called
transitivity surface area, one encounters distinct physical situations
depending on the character of the gradient of $K$. This function, in a
vacuum, electrovacuum or massless scalar case, due to the vanishing of the
two-trace in $t,z$ of the stress tensor, is necessarily a solution of the
wave equation of the form:

\be
\ddot{K}(t,z)-K''(t,z)=0.\label{wave}
\ee

The solutions $K=t$, $K=\sinh{t}\sinh{z}$ or $K=\sin{t}\sin{z}$ are often
used in the studies of anisotropic and inhomogeneous cosmologies and in
colliding wave solutions. The case $K=z\equiv \rho$ corresponds to
Einstein-Rosen cylindrical waves etc. For all these cases the general
solutions of the Klein-Gordon Eq. (\ref{KG1}) are known and well
understood. For example, the different modes of the solution for the case
$K=t$ can be written as

\bea
\Phi = \beta\log{t}+ {\cal L}\{A_{\omega}\cos[\omega(z+z_0)]J_{0}(\omega
t)\} \nonumber \\ + {\cal L} \{B_{\omega}\cos[\omega(z+z_0)]N_{0}(\omega
t)\} + \sum_{i} d_{i}{\rm arc}\cosh\left(\frac{z+z_{i}}{ t}\right),
\nonumber
\eea
where ${\cal L}$ indicates linear combinations of the terms in curly
brackets, $\omega$ can have a discrete or continuous spectrum and $\beta,
A_{\omega}, B_{\omega}, d_{i}$ are constants. The ${\rm arc}\cosh $ terms
are somewhat special in the sense that these can not be written as
Fourier-Bessel integrals \cite{yo} and are often referred to as
gravitational solitons \cite{solitons} due to the relation to the inverse
scattering technique where these terms usually pop up.  In a more general
case, when the gradient of $K$ may vary from point to point and we are
interested in either cosmologies with $S^{3}$ topology of spatial
sections, as it happens in the case, for example, of Bianchi IX models,
Gowdy models or in a colliding wave problem, the function $K$ may be taken
as $K\sim \sin{t} \sin{z}$, and the general solution of Eq. (\ref{KG1})
can be expanded in Legendre polynomials of the first and second kind,

\bea
\Phi &=& \alpha_{1}\log\left|\tan{\frac{t}{2}}\right| + \alpha_{2}
\log\left|\tan \frac{z}{2}\right|+ \alpha_{3}\log|\sin{t}\sin{z}| \nonumber \\
&& + \sum_{\ell=0}^{\infty}\left[ A_{\ell}P_{\ell}(\cos t)+  B_{\ell}
Q_{\ell}(\cos t)\right]
\left[C_{\ell}P_{\ell}(\cos t)+D_{\ell}Q_{\ell}(\cos t) \right],
\nonumber\eea
where $\alpha_i, A_{\ell}, B_{\ell}, C_{\ell}, D_{\ell}$ are constants.

\subsection{A simple example with $K=t$}

In cosmology one is often interested in the solutions for which the
gradient of the transitivity surface $K$ is globally timelike because most
of the homogeneous models have this property.  In this case one may choose
$K=t$ as a solution of Eq. (\ref{wave}). The contribution of the scalar
field to the function $f$ becomes then

\be
\dot{f}(t,z)^{\rm sc}= t\left(\sum_{i=1}^{N}\dot{\varphi}_{i}^{2}+
\sum_{i=1}^{N}\varphi_{i}'^{2}\right),
\ee

\be
f'(t,z)^{\rm sc} = 2t\sum_{i=1}^{N}\dot{\varphi}_{i}\varphi_{i}'.
\ee

As a seed let us take, for example, the vacuum Kasner solution for which
one has:

\be
ds^2 = e^{f(t,z)}(-dt^2 + dz^2)+ \,t\,(e^{p(t,z)}dx^2+ e^{-p(t,z)}dy^2),
\ee
with
\be p= k\log{t}, \qquad f= \frac{k^2-1}{2}\log{t}.
\ee

The scalar field equation becomes now $\ddot{\Phi}+ \dot{\Phi}/t \, -
\Phi'' =0$, and one may take the simplest homogeneous solution as:

\be
\Phi= a\log{t}.
\ee

Choosing the inhomogeneous scalar field solution would have lead to an
inhomogeneous scalar field generalization of the Kasner (Bianchi I) model,
and when lifted to higher dimensions would have produced inhomogeneous
vacuum solutions in higher dimensions.

The corresponding function $f=f^{\rm vac}+\, f^{\rm sc}$  becomes:

\be
f= \frac{k^2+a^2-1}{2}\log{t}.
\ee

The vacuum 5-D solution is then easily obtained using the expressions
above, and has the following synchronous form:

\be
ds^2 =-dt^2+t^{A}dz^2+t^{B}dx^2+t^{C}dy^2+t^{D}dw^2,
\ee
where

\begin{eqnarray}
A&=&\frac{6(k^2+a^2-1)-8 \sqrt{3}a}{3k^2+3a^2+9-4\sqrt{3}a},\quad
B=
\frac{12(k+1-2\sqrt{3}a/3)}{3k^2+3a^2+9-4\sqrt{3}a}, \nonumber\\
C&=&\frac{12(1-k-2\sqrt{3}a/3)}{3k^2+3a^2+9-4\sqrt{3}a},\quad
D=\frac{16\sqrt{3}a}{3k^2+3a^2+9-4\sqrt{3}a}.
\end{eqnarray}

\section{ One Timelike Killing Direction}

We now turn to the case where one of the Killing vectors is timelike. This
situation corresponds to the stationary axially symmetric spacetimes. The
procedure to build higher dimensional solutions is similar to the
previously discussed one but with some minor sign changes, once an adapted
coordinate system is used.

Surprisingly, little is known on scalar field generalizations of axially
symmetric spacetimes.  Basically, this is due to the fact that scalar
field stationary axisymmetric solutions are probably not that exciting.
Unlike in cosmology, where the scalar fields play central role in
inflation, dark matter and dark energy models, the interest in scalar
field generalizations of axially symmetric solutions of the Einstein field
equations is rather scarce. One of the reasons as to why the scalar fields
in axisymmetric spacetimes are of little interest, is because these do not
admit, apart from some very special cases, a perfect fluid description as
in cosmological case, where the scalar field serves as velocity potential
for the fluid. Moreover, the various no-hair theorems \cite{bekenstein1}
exclude scalar field black holes in 4-dimensions. Several specific
solutions, however, in the spherical case, Kerr-type generalizations
\cite{agnese} and the case with conformally coupled scalar fields
\cite{bekenstein} are known. A specific algorithm which converts a
gravitational degree of freedom into a scalar field is also known
\cite{JRW}, but is less adapted for the purposes of this paper.

Our starting point, this time, is the following line element
\cite{belinsky}:

\be
\label{as}
ds^2=e^{\sigma^{\rm vac}}(dr^2+dz^2)+\gamma_{ab}\,dx^a\,dx^b,
\ee
which we take to be a {\em solution} of the vacuum Einstein equations in
$4$ dimensions. The function $\sigma^{\rm vac}$ and $\gamma_{ab}$ are now
functions of $z$ and $r$ alone and $(\phi,t)\equiv(x^{3},x^{0})$. In this
case, as long as the determinant of $\gamma$ is not a constant, we may
take without any lost of generality \cite{belinsky}, 

\be \det \gamma=-r^2.
\ee

We now assume, as in the previous section, that the scalar field
$\Phi\equiv \sum_{i=1}^{N}\varphi_{i}$ is a solution of the following
equation:

\be
\Phi_{rr} + \frac{1}{r}\,\Phi_{r}+\Phi_{zz}=0. \label{KG}
\ee

This equation is easily obtained from the Klein-Gordon equation
(\ref{KG1}) of the previous section with the following change of
coordinates $t\to r,\, z\to iz$ and is a consequence of the formal
relationship between the $G_{2}$ - Generalized Einstein-Rosen class, and
the $G_{2}$ - Stationary Axially Symmetric class metrics.  If the vacuum
solution is globally diagonalizable (the static Weyl case) then the ``Weyl
potential" $U$ which appears as a metric function $e^{U}\,dt^2$ (see
Appendix A), solves exactly the same Eq. (\ref{KG}) as $\Phi$.  The
general solution to the linear Eq. (\ref{KG}) is obtained by considering
the following integral:

\be
\Phi=\int^{\infty}_{\alpha=-\infty}\int^{2\pi}_{\beta=0}
\frac{{F(\alpha})d\alpha
d\beta}{\sqrt{r^2+(z-\alpha)^2+G^{2}(\alpha)-2G^{2}(\alpha)\cos\beta}}.
\ee

It is convenient, nevertheless, in a way analogous to the solutions of Eq.
(\ref{KG1}), to express the solution of the Eq. (\ref{KG}) as a
superposition of the following terms:

\bea
\Phi &=& \beta\log{r}+ {\cal L}\{A_{\omega}\cosh[\omega(z+z_0)]J_{0}(\omega
r)\} \nonumber \\
&&+ {\cal L}
\{B_{\omega}\cosh[\omega(z+z_0)]N_{0}(\omega r)\}
+ \sum_{i} d_{i}{\rm arc}\sinh \left(\frac{z+z_{i}}{r}\right),
\nonumber
\eea
where the ${\rm arc}\sinh$ terms are the Weyl-analogs of the ${\rm
arc}\cosh$ terms, and can be written as:

\be
{\rm arc}\sinh{\frac{z+m}{r}=\log \left[(z+m)+\sqrt{r^2+(z+m)^2}\right]}
-\log{r}.
\ee

The arbitrary constant $m$ is often called a soliton ``pole"  and may be
either real or complex in which case one must take ${\cal R}e \,\left[{\rm
arc}\sinh{\frac{z+m}{r}}\right]$ or ${\cal I}m \,\left[{\rm
arc}\sinh{\frac{z+m}{r}}\right]$ as the solution. If the soliton has a
real pole, then the pole is directly related to the ``rod" structure of
the Weyl solutions. In fact it is an interesting way to exactly perturb
the solitonic solutions by allowing the poles $m$ to ``catch" some
imaginary part $m+ i\epsilon$, see for example \cite{Feinst-Charach}.

\subsection{Coupled Einstein-scalar field equations}

It is simple to show that the solution to the coupled Einstein-scalar
field equations (see Appendix A), independently whether static or
stationary, can be obtained by keeping the transverse part characterized by
the metric function $\gamma_{ab}$ without a change, but with the
longitudinal function $\sigma(r,z)^{\rm vac}$ replaced by

\be
\sigma(r,z)^{\rm vac} \longrightarrow \sigma(r,z)^{\rm vac} +
\sigma(r,z)^{\rm sc}.
\ee

The function $\sigma(r,z)^{\rm sc}$ is then solved by quadratures from:
\be
\sigma^{\rm sc}_{r}=r\left[
\sum_{i=1}^{N}\varphi_{ir}^{2}-
\sum_{i=1}^{N}\varphi_{iz}^{2}\right], \label{fr}
\ee

\be
\sigma^{\rm sc}_{z} =2\,r\left[\sum_{i=1}^{N}{\varphi}_{ir}\varphi_{iz}
\right]. \label{fz}
\ee

To find the above expressions one can either, directly work with the
Einstein equations as in \cite{letelier}, or put in the previous
all-spacelike Killing vector case $K=t$ and perform formally the following
coordinate transformation:

\be
t\mapsto r,\, z\mapsto iz,\, x\mapsto i\theta,\, y\mapsto t.
\ee

One should also perform a global signature change after Wick rotating the
solution if one desires to maintain the same signature. Equation-wise, but
not solution-wise, there is a one-to-one correspondence between the case
of all spacelike Killing vectors and the case when one of the Killing
directions is timelike. Some solutions do not have stationary analogue and
vice versa, especially, when the two pertinent Killing vectors are not
hypersurface orthogonal. In the diagonal case, however, all the solutions
can be formally ``copied" from one case to another.

From here one may now use the previous lifting expressions for
$\psi_i={\cal D}_{ij}\varphi_{j}$ and the Eq. (\ref{higher}) to construct
the higher dimensional solutions. Before proceeding any further, some
remarks are in order. In the static axially symmetric case, the $4$
dimensional solutions of the vacuum Einstein equations depend just on one
function $U$-the Weyl potential, which as mentioned above, solves Eq.
(\ref{KG}) with $\Phi \to U$ in vacuum or in the scalar case. In the
higher dimensional case there may be an extra scalar degree of freedom for
each extra dimension. In the stationary case, one should allow for an
additional rotational degree of freedom, but, at any rate, we assume that
the vacuum 4 dimensional solutions are already given and do not enter into
their generation-there is little to add on what is already known in this
field \cite{Kramer}.

A separate remark is about what kind of solutions are ``interesting" for
the scalar field in the higher dimensional generalization of axially
symmetric solutions. We believe that those of interest happen to be the
same solutions as the ones producing interesting Weyl potentials (the
solitonic terms):  $\log{[(z+a)+\sqrt{(z+a)^2+r^2}]}\equiv \log{r}+ {\rm
arc} \sinh{[(z+a)/r]}$,\, $\log{r}$ and their linear combinations. Note
that the linear Eq. (\ref{KG}) allows solutions obtained by reflection ($z
\rightarrow -z$), by shift ($z \rightarrow z+a$) as well as by multiplying
the solution by an arbitrary constant $A$. Moreover, if a complex function
$\Phi$ is a solution of Eq. (\ref{KG}), then both the real and the
imaginary parts of $\Phi$ are also solutions.

\section{The 5-D case}

Let us now specialize to a 5-dimensional case. The above described
procedure of constructing a 5-D generalization of the axially symmetric
solutions may be put in a more compact statement. Consider we have a
vacuum solution to the Einstein field equations in 4-dimensions of the
form:

\be
ds^2= -e^{U} (dt+Ad\phi)^2+e^{-U}r^2d\phi^2+e^{\sigma^{\rm
vac}}(dr^2+dz^2).
\label{axi4} \ee

The  following line element
\be
ds^2= -e^{U-\frac{2}{\sqrt{3}} \Phi} (dt+Ad\phi)^2+e^{-U-\frac{2}{\sqrt{3}}
\Phi}r^2d\phi^2
+e^{\sigma^{\rm vac}+\sigma^{\rm sc}-\frac{2}{\sqrt{3}}\Phi}
(dr^2+dz^2)+ e^{\frac{4}{\sqrt{3}}\Phi} dw^2,
\label{axi5}
\ee
is a vacuum solution of the 5-D Einstein equations, provided $\Phi$
is a solution of the Eq. (\ref{KG}) and $\sigma^{\rm sc}$ is given by

\be
\sigma^{\rm sc}_{r}=r\left(
\Phi_{r}^{2}-
\Phi_{z}^{2}\right), \label{fr1}
\ee

\be
\sigma^{\rm sc}_{z} =2\,r\Phi_{r}\Phi_{z}. \label{fz2}
\ee

Thus we see that the construction of the vacuum generalizations of the 4-D
axially symmetric solutions is reduced to simple algebra.

\subsection{Coping the Weyl analogues}

Before aplying the algorithm to generate the stationary solutions we may
start by copying the already known solutions obtained with the two
space-like Killing vectors into stationary solutions. Obviously most of
these ``copies" will not have interesting physical properties.

To exemplify such a direct copying of solutions from cosmology to their
Weyl analogue, we consider the open scalar Friedmann-Robertson-Walker
(FRW) universe. To construct a scalar field FRW cosmology with open
spatial section one starts with the following solution to the vacuum
Einstein equations:

\begin{eqnarray}
ds^2_{\rm vac}&=&(\sinh{2t})^{-\half}(\cosh{4t}-\cosh{4z})^{\frac{3}{4}}
(-dt^2+dz^2) \nonumber\\
&& + \half\sinh{2t}\sinh{2z}\left(\tanh{z}\,dx^2+
{\rm cotanh\,}z\,dy^2\right),
\end{eqnarray}
and ``dresses" it with the scalar field:

\be
\Phi= \frac{\sqrt{3}}{2}\log{\tanh{t}}.
\ee

Immediately one gets a solution which describes an isotropic homogeneous
universe with spatial sections of negative curvature \cite{feinmavm},

\be
ds^2= \sinh{2t}(-dt^2+dz^2)
+ \frac{1}{2} \sinh{2t}\sinh{2z}\left(\tanh{z}\,dx^2+
{\rm cotanh\,}z\,dy^2\right).
\ee

To pass to  Weyl coordinates, we first choose:
\be
T=\sinh{2t}\sinh{2z}, \quad Z=\cosh{2t}\cosh{2z},
\ee
and express
\be
\log({\tanh{t}}) =\frac{1}{2}
\rm{arc}\cosh
\left({\frac{1-Z}{T}}\right)+\frac{1}{2}\rm{arc}\cosh
\left({\frac{1+Z}{T}}\right), \ee
and
\be
\log({\tanh{z}}) =\frac{1}{2}
\rm{arc}\cosh
\left({\frac{1-Z}{T}} \right) -\frac{1}{2}\rm{arc}\cosh
\left({\frac{1+Z}{T}}\right). \ee

After some algebra we find that the Weyl potential and the scalar field
for the analogue of the open FRW universe are:

\be
e^{U}=\log{r}+\frac{1}{2}\log \left[
{\frac{\sqrt{r^2+(z-1)^2}+(z-1)}
{\sqrt{r^2+(z+1)^2}+(z+1)}} \right],
\ee

\be
\Phi=\frac{1}{2}
\log \left\{
{\left[\sqrt{r^2+(z-1)^2}+(1-z)\right]\left[\sqrt{r^2+(z+1)^2}
+(z+1)\right]} \right \},
\ee
for the metric given in \cite{feinmavm}. Physically, the Weyl analogues of
the open FRW universe have nothing to do with the original solution and
probably have little relevance as static solutions. We have presented this
here only to exemplify the procedure of ``copying".

\subsection{ Asymptotic Flatness}

Dealing with the stationary axisymmetric solutions, one often imposes the
asymptotically flat behavior of the line element away from the axis
($r>>z$). In 5 dimensions, this behavior translates into: $g_{ww}\propto
\sqrt{z^2+r^2}+z$, $g_{tt}\propto -1$ and $g_{\phi\phi}/r^2 \propto
g_{ww}^{-1} = \sqrt{z^2+r^2}-z$. Therefore, to build asymptotically flat
solutions we have that if the scalar field $\Phi$ is given by

\be
\Phi=\sum_{i=1}^{N} a_{i}{\varphi}_{i},
\ee
where

\be
\varphi_{i}=\log{\left[(m_{i}-z)+\sqrt{(m_{i}-z)^2+r^2}\right]},
\ee
one must have ($g_{ww}\propto \sqrt{z^2+r^2}+z$)

\be
\frac{4}{\sqrt{3}}\sum_{i=1}^{N} a_{i}=1.
\label{flatness1}
\ee

In the static case if we take the solution for $U$ of the form

\be
U=\sum_{i=1}^{N} b_{i}{V}_{i},
\ee
where
\be
V_{i}=\log{\left[(n_{i}-z)+\sqrt{(n_{i}-z)^2+r^2}\right]},
\ee

to get the asymptotically flat solutions we must impose ($g_{tt}\propto
-1$):

\be
-\frac{2}{\sqrt{3}}\sum_{i=1}^{N} a_{i}+ \sum_{i=1}^{N}
b_{i}=0,\qquad\Longrightarrow \sum_{i=1}^{N} b_{i}=\frac{1}{2},
\ee
the last condition then ($g_{\phi\phi}/r^2 \propto g_{ww}^{-1}$) is
trivially satisfied.

\subsection{The method in terms of the rod structure}

As mentioned before, the interesting Weyl potentials are those associated
with the solitonic terms or ``rods" in the z-axis \cite{bonnor}.
Considering these terms, the generating method can be described as adding
up a source to the fifth dimension. Given the form of the new metric by
(\ref{axi5}), in fact, we are also ``subtracting" the same source from the
other two Killing directions, thus ``compensating" the extra sources that
we have introduced to the system.

We can add either a finite rod of length $(a_0-a)$, or the interval
$(a,a_0)$, by choosing

\begin{equation}
\Phi = \log { \left[ \frac{ \sqrt{r^2+(z+a)^2}-(z+a)}{
\sqrt{r^2+(z-a_0)^2}-(z-a_0)} \right]},
\end{equation}
or a semi-infinite rod $(a, \infty)$ ($(- \infty,-a)$ taking
the lower sign) with

\begin{equation}
\Phi = \log { \{ \sqrt{r^2+(z \mp a)^2}\mp(z \mp a) \} }.
\end{equation}

\section{Trapped surfaces and horizons
of the generated solutions}

In $N>4$ is not easy to figure out topological features of spacetime; to
extract interesting information one needs invariant objects. One of the
most interesting properties to study in these spacetimes is the trapness
of 2-dimensional surfaces. These are imbedded spatial surfaces such that
any portion of them has a decreasing area along any future evolution
direction. A practical way to study the trapped surfaces and locate
horizons was introduced in \cite{Seno} through evaluating a certain scalar
$\kappa$. The sign of this scalar defining the trapping of a surface $S$.
We shall analyze the effect of introducing rods in the 5-D generated
spaces.

For completeness we include some steps in the construction of such scalar
$\kappa$ introduced in \cite{Seno}. Let us consider the line element

\be
ds^2=g_{ab}dx^a dx^b +2 g_{aA}dx^a dx^A+ g_{AB}dx^A dx^B,
\ee
and a family of $(D-2)$-dimensional spacelike surfaces $S_{X^a}$ with
intrinsic coordinates $\{ \lambda_A \}, A,B,..=2,...,D-1$, imbedded into
the spacetime. There are fixed coordinates, $\{x^a=X^a\}, \quad a,b=0,1,
...,X^a$, while $x^{A}$ denote the local coordinates on the surface. $G=
\sqrt{{\rm det}g_{AB}}= e^{U}$ gives the canonical $(D-2)$ volume element
of the surfaces $S_{X^a}$. Introducing $H_{\mu}= \delta^{a}_{\mu} (U_{,a}-
\nabla \cdot g_{a})$, $g_{a} =: g_{aA} dx^A$, where the divergence
operator acts on vectors at $S_{X^a}$, the invariant $\kappa$ is defined
by

\begin{equation}
\kappa_{\{X^a\}}= -g^{bc}H_bH_c |_{X^a}.
\end{equation}

The hypersurfaces $\cal{H}$, defined locally by the vanishing of $\kappa$,
are the so called $S_{X^a}$-horizons \cite{Seno}, and coincide in many
instances with the classical horizons.

In what follows we have found it convenient to work in prolate spheroidal
coordinates $(x,y)$ which make the algebra much easier. These are related
to the Weyl coordinates by

\begin{equation}
r= \alpha \sqrt{(x^2-1)(1-y^2)}, \quad z= \alpha xy,
\end{equation}
with ranges $x \ge 1$ and $-1 \le y \le 1$.

In prolate spheroidal coordinates the generated spacetimes (\ref{axi5})
take the following form:
\be
ds^2=e^{- \frac{2 \Phi}{ \sqrt{3}}}
\left\{ e^{\sigma^{\rm vac} +\sigma^{\rm sc}}\right.
\alpha^2(x^2-y^2) \left( \frac{dx^2}{x^2-1} + \frac{dy^2}{1-y^2}\right)
+ \gamma_{ab}dx^a dx^b \left. \right\} + e^{\frac{4 \Phi}{ \sqrt{3}}}
d \omega^2,
\ee
where $a, b= t, \phi$. We shall consider spacelike surfaces with $t=$const
and $x=$const.  Taking $\{ x^a \}= \{t,x \}$ and
$\{ x^A \}= \{y, \phi, \omega \}$, the scalar
$\kappa_{\{t,x\}}$ is given by

\begin{equation}
\kappa_{\{t,x\}}= - e^{\frac{2 \Phi}{ \sqrt{3}}- \sigma^{\rm sc}
- \sigma^{\rm vac}}
\frac{(x^2-1)}{4 \alpha^2 (x^2-y^2)^3 {\gamma_{\phi \phi}}^2}
{U}_{x}^2,
\end{equation}
with
\begin{equation}
{ U}_{x}= \gamma_{\phi \phi}[(x^2-y^2)(\sigma^{\rm vac}_{x}
+\sigma^{\rm sc}_{x})+2x]+(x^2-y^2)\gamma_{\phi \phi,x}.
\end{equation}

The invariant $\kappa$, as compared with that one of the seed, is modified
by the factor $e^{\frac{2 \Phi}{ \sqrt{3}}-\sigma^{\rm sc}}$. The
inclusion of a semi-infinite rod $(- \infty, -a_0)$ produces therefore

\begin{equation}
e^{\frac{2 \Phi}{ \sqrt{3}}-\sigma^{\rm sc}} =
(x+y)^{4A^2}a_0^{\frac{2A}{ \sqrt{3}}}
[(x+1)(1+y)]^{\frac{2A}{\sqrt{3}}-4A^2},
\end{equation}
while with the finite rod $(a,a_0)$, the modification
corresponds to

\begin{equation}
e^{\frac{2 \Phi}{ \sqrt{3}}-\sigma^{\rm sc}} =
\left( \frac{a}{a_0} \right)^{\frac{2A}{ \sqrt{3}}}(x^2-y^2)^{4A^2}
\frac{(x-1)^{\frac{2A}{\sqrt{3}}-4A^2}}{(x+1)^{\frac{2A}
{\sqrt{3}}+4A^2}}.
\label{factor}
\end{equation}

Hence, using this method new horizons or singularities arise, depending on
the value of the exponent $2A/ \sqrt{3}-4A^2$. For instance, in the case
of a finite rod, if $2A/ \sqrt{3}-4A^2 > 0$ in Eq. (\ref{factor}),
$\kappa$ vanishes at $x=1$ and there is a horizon. While if $2A/
\sqrt{3}-4A^2 < 0$, $\kappa$ diverges at the same point.

\section{Generating static 5-D solutions}

We now illustrate how the method works by generating some static 5-D
solutions by taking Minkowski 4-D and the Schwarzschild 4-D black hole as
seeds.

\subsection{Generating 5-D solutions from  Minkowski seed.}

We first consider Minkowski 4-D spacetime in the Rindler coordinates of
uniformly accelerated observers \cite{emparan}, \cite{Kramer}

\begin{eqnarray}
ds_{M4}^2&=&- e^{2U_1} dt^2+e^{2U_2} d \phi^2 + e^{\sigma^{\rm
vac}}(dr^2+dz^2),
\nonumber\\
U_1&=& \frac{1}{2} \log [-a+z + \sqrt{(-a+z)^2+r^2}] + {\rm const},
\nonumber\\
U_2&=& \frac{1}{2} \log [a-z + \sqrt{(a-z)^2+r^2}] + {\rm const},
\nonumber\\
\sigma^{\rm vac} &=& - \log[r^2+(a-z)^2].
\end{eqnarray}

\begin{figure}
\centering
\includegraphics[width=8.6cm,height=4cm]{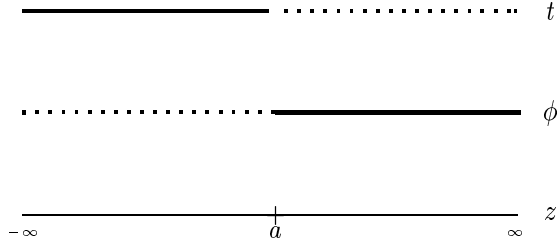}
\caption{\label{fig1}
Rod structure of Minkowski 4-D spacetime}
\end{figure}

The corresponding rod structure consists of a semi-infinite rod $(-
\infty,a)$ in the $\partial_{t}$ direction and a semi-infinite rod $(a,
\infty)$ in the $\partial_{\phi}$ direction, as is shown in Fig.
\ref{fig1}.

Now, consider adding up a semi-infinite rod $(- \infty, -a_0)$ in a fifth
dimension using $\Phi = A \log [a_0+z + \sqrt{(a_0+z)^2+r^2}]$, where $A$
is a constant characterizing the scalar charge and $a_0$ defines a new
interval on the z-axis. The method produces a 5-D solution given by

\be
ds^2_{M5}= e^{- \frac{2}{\sqrt{3}} \Phi + \sigma^{\rm vac}
+ \sigma^{\rm sc}}( dr^2 + dz^2)
+e^{-\frac{2}{\sqrt{3}} \Phi}[- e^{2U_1} dt^2+e^{2U_2} d
\phi^2]+e^{\frac{4}{\sqrt{3}}\Phi}d \omega^2.
\label{5dbh}
\ee
with $\sigma^{\rm sc}= 4 A^2 \log \left[ \frac{a_0+z+
\sqrt{r^2+(a_0+z)^2}}{\sqrt{r^2+(a_0+z)^2}} \right]$.

In prolate spheroidal coordinates it becomes

\begin{widetext}
\begin{eqnarray}
ds^2_{M5}&&= a_0^{-\frac{2A}{\sqrt{3}}} \left\{
(x+y)^{1-4A^2}[(1+y)(1+x)]^{4A^2-\frac{2A}{\sqrt{3}}} \right.
\left( \frac{dx^2}{x^2-1}+ \frac{dy^2}{1-y^2} \right)  \nonumber\\
&& + (x+1)^{1-\frac{2A}{\sqrt{3}}}(1+y)^{-\frac{2A}{\sqrt{3}} }(1-y) d
\phi^2
- (1+y)^{1-\frac{2A}{\sqrt{3}}}(1+x)^{-\frac{2A}{\sqrt{3}} }(x-1) d t^2
\nonumber\\
&& +a_0^{\frac{6A}{\sqrt{3}}}[(x+1)(1+y)]^{\frac{4A}{\sqrt{3}}} d \omega^2
\left. \right\}.
\label{5dbh-Mink}
\end{eqnarray}
\end{widetext}

The value of the scalar charge $A$ defines several important features of
the generated spacetime. We illustrate it by analyzing the above generated
solutions for two different values of scalar charge: $A=\frac{
\sqrt{3}}{2} $ and $A=\frac{ \sqrt{3}}{4}$.

\subsubsection{Case with scalar charge $A=\frac{ \sqrt{3}}{2}$}

Choosing scalar charge as $A=\frac{ \sqrt{3}}{2}$, (\ref{5dbh-Mink}) gives
\begin{eqnarray}
ds^2_{M5}&&= a_0^{-1} \left\{
(x+y)^{-2}[(1+y)(1+x)]^{2} \right.
\left( \frac{dx^2}{x^2-1}+ \frac{dy^2}{1-y^2} \right) \nonumber\\
&& + \frac{(1-y)}{(1+y)} d \phi^2 -\frac{(x-1)}{(x+1)} d t^2
\left. \right\}
+ [a_0(x+1)(1+y)]^{2} d \omega^2.
\label{5dbh-MP}
\end{eqnarray}

\begin{figure}
\centering
\includegraphics[width=8.6cm,height=4cm]{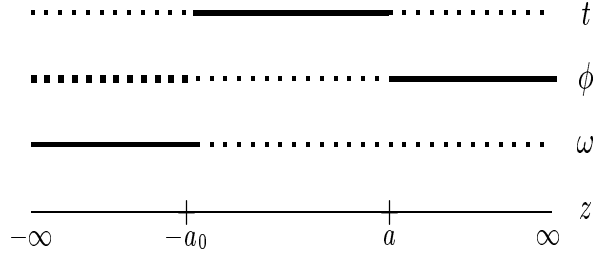}
\caption{\label{fig2}
Rod structure of the solution generated using Minkowski 4-D
and inserting a semi-infinite rod $(- \infty, -a_0)$ in the fifth
dimension with scalar charge $A=\frac{ \sqrt{3}}{2}$.
The bold dotted line along
$\partial_{\phi}$ corresponds to a rod with negative mass density.}
\end{figure}

For $x=1$ ($r=2m$), $g_{tt}=0$ and a horizon is present, while $g_{\phi
\phi}$ diverges at $y=-1$.  The solution is not asymptotically flat as
seen from the behavior of $g_{\omega \omega}$. The corresponding rod
structure is as follows (Fig. \ref{fig2}).

A finite rod $(-a_0, a)$ in the $\partial_{t}$ direction (event horizon),

a semi-infinite rod $(a, \infty)$ in the $\partial_{\phi}$ direction
(axis of rotation of $\phi$),

a semi-infinite rod $(- \infty,-a_0)$  with negative mass density in the
$\partial_{\phi}$ direction and

a semi-infinite rod $( - \infty,-a_0)$ in the $\partial_{w}$ direction.


\begin{figure}
\centering
\includegraphics[width=8.6cm,height=4cm]{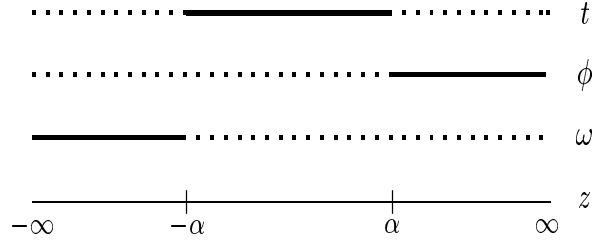}
\caption{\label{fig3}
Rod structure of Myers-Perry static 5-D black hole.}
\end{figure}

The previous solution (\ref{5dbh-MP}) can be compared with the static
Myers-Perry (MPs) black hole \cite{Myers} (Eq. (5.16) without rotation,
$a_1=a_2=0$ in \cite{harmark}), whose line element and rod structure (Fig.
\ref{fig3}) we include here for completeness:

\begin{eqnarray}
ds^2_{MPs}&&= \frac{r_0^{2}}{4}
\left\{2(x+1)\right.
\left( \frac{dx^2}{x^2-1}+ \frac{dy^2}{1-y^2} \right)
+(x+1)(1-y) d \phi^2 \nonumber\\
&& - \frac{4}{r_0^2}\frac{(x-1)}{(x+1)} d t^2
+ (x+1)(1+y) d \omega^2 \left. \right\}.
\label{MP_s}
\end{eqnarray}

Comparing Fig. \ref{fig2} with Fig. \ref{fig3}, note the similarity in rod
structure, it is the same except for the presence of the negative mass
density along the Killing direction $\partial_{\phi}$.

We now consider the spacelike surfaces with $t=$const and $x=$const
($r=$const) for both solutions.
The corresponding scalars $\kappa$ that define the trapping of such surfaces
are

\begin{eqnarray}
\kappa_{MPs}&&=- \frac{9}{2 r_0^2} \frac{(x-1)}{(x+1)^2}=- \frac{9m}{2
r_0^2}\frac{(r-2m)}{r^2} ,
\label{kapp-MPs}\\
\kappa_{M5}&&=- a_0 \frac{(x-1)(x+2y-1)^2}{(x+1)^{3}(1+y)^{2}}
\nonumber\\
&&= - a_0 \frac{(r-2m)[r-2m(1- \cos{\theta})]^2}{r^{3}(1+
\cos{\theta})^{2}}.
\label{kapp-M5}
\end{eqnarray}

\begin{figure}
\centering
\includegraphics[width=7cm,height=6.5cm]{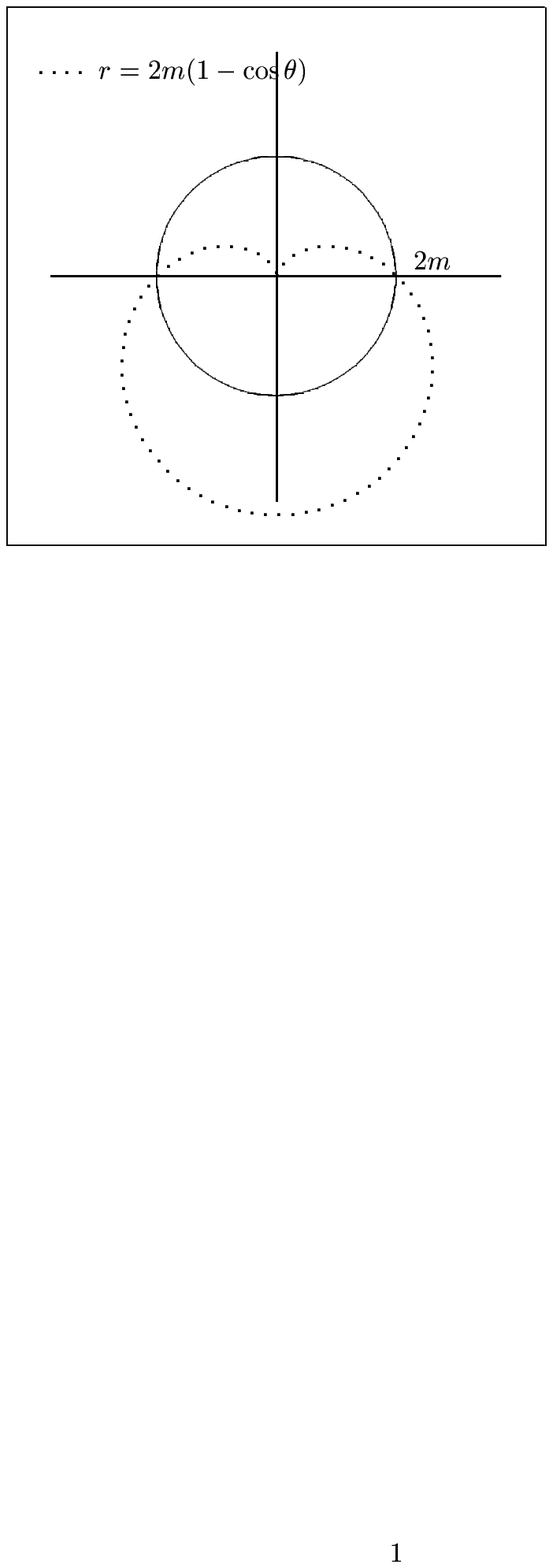}
\caption{\label{fig4}
Squeme of the horizons, for a fixed angle $\phi$
and sweeping $\theta$ for the 5-D solution generated from Minkowski and
adding the rod $(- \infty, -a_0)$ with $A=\frac{\sqrt{3}}{2}$.
The solid circle represents the trapped surface $r=2m$,
while the dotted curve corresponds to $r=2m(1- \cos{\theta})$.}
\end{figure}

\begin{figure}
\centering
\includegraphics{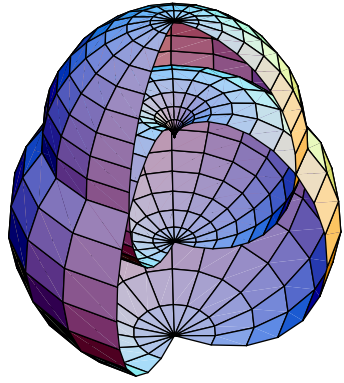}
\caption{\label{fig5}
Marginally trapped surfaces of Minkowski 5-D, generated with a Minkowski
4-D seed plus a dilaton field $\Phi= \frac{ \sqrt{3}}{2}
\log[a_0(x+1)(1+y)]$, these surfaces are the rotated slices of Fig. 
\ref{fig4}.}
\end{figure}

Both scalars exhibit singularity at $r=0$ as well as the horizon at
$r=2m$. Note, however that the solution (\ref{5dbh-MP}) presents an
additional marginally trapped surface ($\kappa=0$) defined by $r=2m(1-
\cos{\theta})$. The scalar $\kappa$ also becomes singular at $\theta=
\pi$. The profiles of the marginally trapped surfaces are shown in Fig.
\ref{fig4} and then rotated in Fig. \ref{fig5}.


\subsubsection{Case $A= \frac{ \sqrt{3}}{4}$}

Another interesting case occurs if we choose $A=\frac{ \sqrt{3}}{4}$ in
(\ref{5dbh-Mink}). The generated solution then becomes

\begin{eqnarray}
ds^2_{M5}&&= a_0^{- \frac{1}{2}} \left\{
[(x+y)(1+y)(1+x)]^{\frac{1}{4}} \right.
\left( \frac{dx^2}{x^2-1}+ \frac{dy^2}{1-y^2} \right) \nonumber\\
&& + \frac{(1-y)(x+1)^{\frac{1}{2}}}{(1+y)^{\frac{1}{2}}} d \phi^2
-\frac{(x-1)(1+y)^{\frac{1}{2}}}{(x+1)^{\frac{1}{2}}} d t^2
\left. \right\} + a_0(x+1)(1+y) d \omega^2.
\label{5dbh-MP2}
\end{eqnarray}

\begin{figure}
\centering
\includegraphics[width=8.6cm,height=4cm]{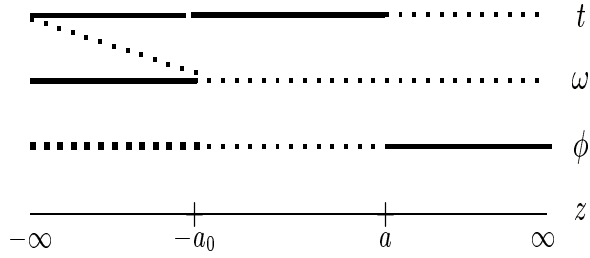}
\caption{\label{fig6}
Rod structure for the 5-D solution generated from Minkowski and
adding the rod $(- \infty, -a_0)$ with $A=\frac{ \sqrt{3}}{4}$.
The dotted line between $t$ and $\omega$ indicates that the
rod $(- \infty, -a_0)$ has components in both directions,
$\partial_{t}$ and $\partial_{\omega}$.}
\end{figure}

This time the $g_{\omega \omega}$ component satisfies the condition for
asymptotic flatness Eq. (\ref{flatness1}), but $g_{tt} \simeq
x^{\frac{1}{2}}$ does not and the solution is not globally asymptotically
flat. The rod structure is shown in Fig. \ref{fig6}.


New marginally trapped surfaces also arise in this case. The scalar
$\kappa_{\{t,r\}}$ that characterizes the trapping of a spacelike surface
of constant $t$ and $r$ for (\ref{5dbh-MP2}) is

\begin{equation}
\kappa_{\{ t,r \}}= - \sqrt{a_0}m^{\frac{3}{4}}
\frac{(r-2m)[r- \frac{7}{8}m(1-\cos{\theta})]^2}
{(1+\cos{\theta})^{\frac{1}{4}}[r-m(1-
\cos{\theta})]^{\frac{9}{4}}r^{\frac{5}{4}}}.
\label{k_M2}
\end{equation}
From Eq. (\ref{k_M2}) we learn that the solution has two marginally
trapped surfaces, one situated at the Schwarzschild horizon, $r=2m$, while
the second apparent horizon is the surface defined by $r=
\frac{7}{8}m(1-\cos{\theta})$, with $r_{\rm max}= \frac{7}{4}m$. This
second surface remains hidden inside $r=2m$.

\subsection{Starting with a Schwarzschild seed.}

\begin{figure}
\centering
\includegraphics[width=8.6cm,height=4cm]{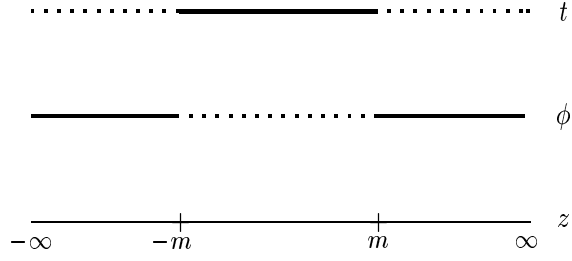}
\caption{\label{fig7}
Rod structure of Schwarzschild 4-D black hole with mass $2m$.}
\end{figure}

Consider now Schwarzschild 4-D line element in prolate spheroidal
coordinates,

\be
ds^2= m^2(x+1)^2 \left( \frac{dx^2}{x^2-1}+ \frac{dy^2}{1-y^2}
\right)
+ m^2(1-y^2)(x+1)^2 d \phi^2- \frac{(x-1)}{(x+1)}dt^2.
\label{Schw-4d}
\ee
The rod structure of this solution is as follows: a finite rod of length
$2m$ in the timelike direction, $\partial_t$, and two semi-infinite rods
$(- \infty, -m), (m, \infty)$ in the spacelike direction
$\partial_{\phi}$, as shows Fig. \ref{fig7}.

\begin{figure}
\centering
\includegraphics[width=8.6cm,height=4cm]{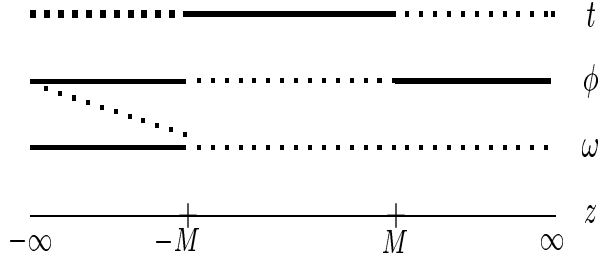}
\caption{\label{fig8}
Rod structure of the static 5-D solution generated from Schwarzschild
spacetime by adding a semi-infinite rod $(- \infty,-m)$ with $\Phi=\frac{
\sqrt{3}}{4} \log[m(x+1)(1+y)]$. The bold dotted line indicates negative
mass density in the interval $(- \infty, -m)$; the dotted line between
${\phi}$ and ${\omega}$ symbolizes that the semi-infinite rod $(- \infty,
-m)$ has components in both directions $\partial_{\phi}$ and
$\partial_{\omega}$.}
\end{figure}

Adding up the semi-infinite rod $(- \infty, -m)$ by taking $\Phi= \frac{
\sqrt{3}}{4} \log[m(x+1)(1+y)]$ and lifting to 5-D, results in the
following spacetime

\begin{eqnarray}
ds^2_{5}&&= m^{\frac{3}{2}}
\left\{(x+1)^{\frac{9}{4}}
\frac{(1+y)^{\frac{1}{4}}}{(x+y)^{\frac{3}{4}}}\right.
\left( \frac{dx^2}{x^2-1}+ \frac{dy^2}{1-y^2} \right)
+(x+1)^{\frac{3}{2}}(1-y)(1+y)^{\frac{1}{2}} d \phi^2 \nonumber\\
&&- \frac{1}{m^2}\frac{(x-1)}{(x+1)^{\frac{3}{2}}} \frac{ d
t^2}{(1+y)^{\frac{1}{2}}}\left. \right\}
+ m(x+1)(1+y) d \omega^2.
\label{Schw-5d}
\end{eqnarray}
This procedure introduces a singularity in $g_{tt}$ due to the factor
$(1+y)^{-\frac{1}{2}}$. The corresponding rod structure is shown in Fig.
\ref{fig8}.


Several interesting properties of the spacetime (\ref{Schw-5d}) are worth
to point out: the metric is asymptotically flat, it has a horizon at $x=1$
and $g_{\phi \phi}$ is finite in all the domain of $x$ and $y$. There is a
divergence in $g_{tt}$ at $y=-1$.

The spacetime possesses, besides the surface $r=2m$, another marginally
trapped surface that deforms the spherical symmetry of the horizon. To see
it we calculate the scalar $\kappa_{\{t,r\}}$ for (\ref{Schw-5d}),

\bea
\kappa_{Schw-5D}&&=- \frac{1}{4^3m^{\frac{3}{2}}}
\frac{(x-1)(16x+19y-3)^2}
{(x+1)^{\frac{13}{4}}(1+y)^{\frac{1}{4}}(x+y)^{\frac{5}{4}}} \nonumber\\
&&=-4 \frac{(r-2m)[r- \frac{19}{16}m(1-\cos{\theta})]^2}
{(1+ \cos{\theta})^{\frac{1}{4}} r^{\frac{13}{4}}[r-m(1-
\cos{\theta})]^{\frac{5}{4}}}.
\label{kapp-Schw}
\eea

\begin{figure}
\centering
\includegraphics[width=7cm,height=6.5cm]{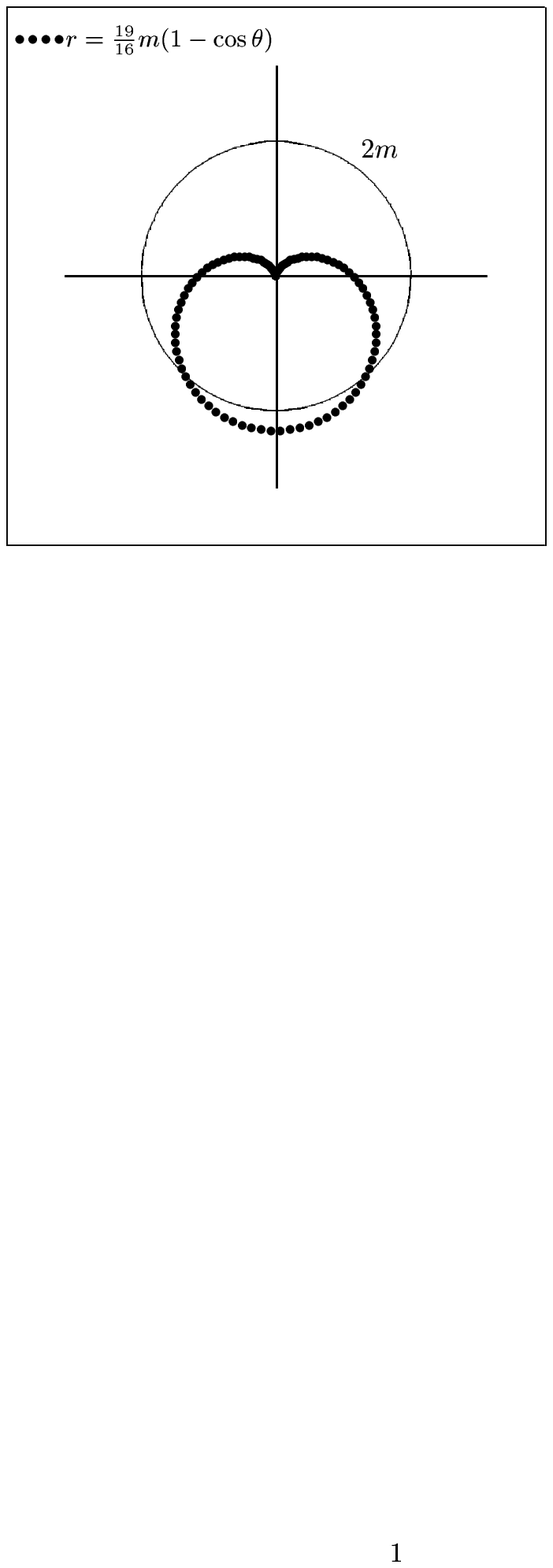}
\caption{\label{fig9}
Squeme of the horizons, for a fixed angle $\phi$
and sweeping $\theta$, of the generated Schwarzschild 5-D black
hole, with $A=\frac{ \sqrt{3}}{4}$; $r_{\rm max}=2.37 m$
at $\theta=\pi$.}
\end{figure}

\begin{figure}[ht]
\centering
\includegraphics{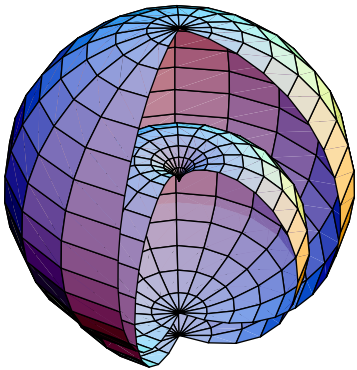}
\caption{
Marginally trapped surfaces of Schwarzschild 5-D, generated with a
Schwarzschild 4-D seed plus a dilaton field $\Phi=\frac{ \sqrt{3}}{4} \log
[(x+1)(1+y)]$; these are the rotated slices of Fig. \ref{fig9}}
\label{fig10} \end{figure}

The vanishing of the scalar $\kappa$ indicates the presence of a
marginally trapped surface, that can be associated to a horizon. For the
solution (\ref{Schw-5d}), $\kappa$ vanishes on two surfaces: $r=2m$, and
$r=\frac{19}{16}m(1-\cos{\theta})$, indicating the distortion of horizon
that we have mentioned before. The new horizon ``intermingles" with
portions of the classical Schwarzschild horizon, presenting for an
external observer a non-spherical horizon of a ``peanut" shape. Also there
is the singularity at $r=0$ where $\kappa$ diverges. Therefore the
solution may be thought of a 5-D black hole distorted by the presence of a
string with a negative mass density, hence deforming its horizon. Slices
of the horizons are shown in Fig. \ref{fig9} and rotated in Fig.
\ref{fig10}. 

\subsection{The static Myers-Perry solution}

We note  in passing that the static Myers-Perry solution (\ref{MP_s})
can be obtained by our lifting method using the appropriated seed:

\begin{eqnarray}
U&=& \frac{1}{4} \log \left[ \frac{(r_{-}+z- \alpha)
(r_{+} + r_{-}- 2\alpha)}
{\alpha (r_{+} + r_{-} + 2\alpha)} \right] \nonumber\\
&=&\frac{1}{4} \log \left[ \frac{(1+y)(x-1)^2}{(x+1)} \right],
\end{eqnarray}
where $r_{\pm}^2=r^2+(z \pm \alpha)^2$. Using $U$ in the line element
(\ref{axi4}) and adding up a semi-infinite rod $(- \infty, -\alpha)$ as in
(\ref{axi5}), with the dilaton

\bea
\Phi&&= \frac{\sqrt{3}}{4} \log[(x+1)(1+y)] \nonumber\\
&&=\frac{\sqrt{3}}{4} \log[\sqrt{r^2+(z+ \alpha)^2}
+z+ \alpha].
\eea

We do not discuss this case further for it was thoroughly done in the
literature.

\subsection{Changing the dilaton}

We again start with the Schwarzschild seed, line element (\ref{Schw-4d})
or (\ref{Schw-4d,Weyl}). Taking the dilaton field as

\be
\Phi=c\,\left\{ {\rm arc}\sinh{\frac{(m+z)}{r}}\,+\, {\rm
arc}\sinh{\frac{(m-z)}{r}} \right\}.
\label{Phi}
\ee
Now, changing to prolate spheroidal coordinates, $(x,y)$,
\bea
x=&(\sqrt{(z+m)^2 +r^2}+\sqrt{(z-m)^2 +r^2})/2m,\\
y=&(\sqrt{(z+m)^2 +r^2}-\sqrt{(z-m)^2 +r^2})/2m,
\eea
the dilaton field (\ref{Phi}) is written as
\be
\Phi=c\, \log
\left(\frac{x-1}{x+1}\right),
\ee
changing now to curvature coordinates $(r, \theta)$:
\be
x=\frac{r}{m}\, -1\, ,\, y=\cos{\theta},
\ee
we find the following 5-D metric:

\bea
ds^2&=&\left(1-\frac{2m}{r} \right)^{- \frac{2c}{\sqrt{3}}}
\left[- \left(1-\frac{2m}{r} \right)^{a}\,dt^{2}+ \right.
\left(1-\frac{2m}{r} \right)^{-a}\,dr^{2} \nonumber\\
&& + \left. \left(1-\frac{2m}{r} \right)^{1-a}\,r^2\,(d\theta^2+
\sin{\theta}^2\,d\phi^2) \right]+
\left(1-\frac{2m}{r} \right)^{\frac{4c}{\sqrt{3}}}\,dw^2,
\eea
here $a=\sqrt{1-4c^2}$. We recognize the metric in square brackets as the
4-D scalar solution derived in \cite{agnese}.

The corresponding scalar $\kappa_{\{t,r\}}$ is given by

\be
\kappa_{\{t,r\}}= - \frac{4}{r^2} \left(1- \frac{m(1+a)}{r} \right)^2
\left(1- \frac{2m}{r} \right)^{a+\frac{2c}{\sqrt{3}}-2},
\label{k_a}
\ee

In general this spacetime does not possess a regular horizon, as can be
seen from (\ref{k_a}) since the exponent $a+\frac{2c}{\sqrt{3}}-2 <0$.
There is a marginally trapped surface at $r=m(1+a)$ that is always hidden
inside the singular surface $r=2m$ since $a<1$. However, an interesting
situation occurs when $c=\sqrt{3}/4$ when the metric becomes:

\be
ds^2=-dt^{2}+ \left( 1-\frac{2m}{r}
\right)^{-1}\,dr^{2}\,+\,r^2\,(d\theta^2+
\sin^2{\theta}\,d\phi^2)
+ \left(1-\frac{2m}{r} \right)\,dw^2.
\label{naked_s}
\ee

This spacetime is asymptotically flat. Performing the Wick rotation $t
\mapsto it, \quad \omega \mapsto i \omega$ we finish with the black string
$S^2 \times R$.

\section{Generation of 5-D stationary solutions}

\begin{figure}
\centering
\includegraphics[width=8.6cm,height=4cm]{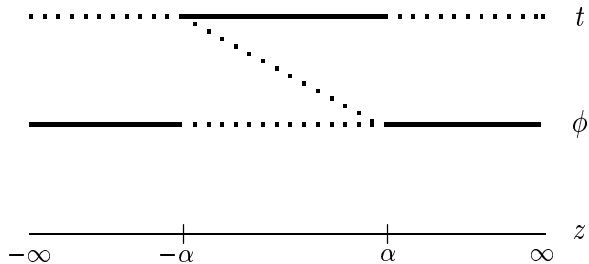}
\caption{\label{fig11}
Rod structure of the stationary Kerr black hole;
the dotted line that intersects $\partial_t$ and $\partial_{\phi}$
is intended to indicate that the orientation of the finite  rod
$(- \alpha, \alpha)$ has one component along $\partial_t$ and other along
$\partial_{\phi}$.}
\end{figure}

According to the above outlined method, to construct nonstatic solutions
one must start with vacuum stationary solutions, since the method does not
introduce non diagonal elements into the generated 5-D metric. The
simplest stationary seed is the Kerr solution, whose line element in
Boyer-Lindquist coordinates $(t, r, \theta, \phi)$ is:

\begin{eqnarray}
ds_{Kerr}^2 &=& - \frac{\Delta-a^2 \sin^2{\theta}}{\Sigma}dt^2 -2a
\sin^2{\theta}
\frac{r^2+a^2- \Delta}{\Sigma}dt d\phi \nonumber\\
&& + \frac{(r^2 +a^2)^2- \Delta a^2 \sin^2{\theta}}{\Sigma}
\sin^2{\theta} d \phi^2+ \Sigma \left( \frac{dr^2}{\Delta}+ d \theta^2
\right),
\end{eqnarray}
where $\Delta= r^2- 2mr +a^2$ and $\Sigma= r^2+a^2 \cos^2{\theta}$.  The
corresponding rod structure is shown in Fig. \ref{fig11}.

Using Kerr solution as seed we construct deformations to the Myers-Perry
5-D rotating black hole. The generated solutions present a rod structure
very close to the original undeformed one, except for some new
singularities that can be avoided using a different seed. In what follows
we explore two cases: the Kerr seed plus a semi-infinite rod $(- \infty, -
\alpha)$ with scalar charge $A= \frac{ \sqrt{3}}{2}$ and Kerr seed again
with an extra distinct semi-infinite rod $(a_0, \infty)$ with the same
scalar charge.

\subsection{Inserting a semi-infinite rod $(- \infty, - \alpha)$}

Lifting the Kerr solution to a fifth dimension by adding the semi-infinite
rod $(- \infty, - \alpha)$, corresponding to $\Phi=\frac{ \sqrt{3}}{2}
\log[ \alpha (x+1)(1+y)]$, we obtain the following line element,

\begin{eqnarray}
ds^2&&= e^{\sigma} \left( \frac{dx^2}{x^2-1}+ \frac{dy^2}{1-y^2}  \right)
+ g_{ij}dx^i dx^j, \quad i,j=t, \phi, \omega,
\nonumber\\
e^{\sigma}&&= \frac{m^2}{ \alpha}\frac{(x+1)^2(1+y)^2}{(x+y)^3}
[(1+px)^2+q^2y^2], \nonumber\\
g_{tt}&&=- \frac{1}{\alpha(x+1)(1+y)}
\frac{(p^2x^2+q^2y^2-1)}{[(1+px)^2+q^2y^2]}, \nonumber\\
g_{t \phi}&&=- \frac{2a(1-y)}{ \alpha(x+1)}
\frac{(1+xp)}{[(1+px)^2+q^2y^2]}, \nonumber\\
g_{\phi \phi}&&=- \frac{(1-y)}{ \alpha(x+1)}
\frac{ \{4a^2(1-y^2)(1+xp)^2-
\alpha^2(x^2-1)[(1+px)^2+q^2y^2]^2
\}}{[(1+px)^2+q^2y^2][p^2x^2+q^2y^2-1]},
\nonumber\\
g_{\omega  \omega}&&=[\alpha(x+1)(1+y)]^2,
\label{Kerr1-5d}
\end{eqnarray}
where the parametrization is $p=\alpha/m= \sqrt{m^2-a^2}/M$, $q=a/m$ and
$p^2+q^2=1$, $a$ and $m$ stand for the acceleration and mass,
respectively. From the inspection of $g_{\omega \omega}$ it is apparent
that the solution is not asymptotically flat and an extra source is
introduced. The rod structure of (\ref{Kerr1-5d}) is analyzed in detail in
Appendix B and it is shown in Fig. \ref{fig12}. The rod structure
resembles the one of Myers-Perry rotating black hole \cite{Myers}, except
for the divergence of $g_{tt}$ in the interval $(- \infty, - \alpha)$.  
In spite of the similarity in the rod structure, the corresponding metric
functions are different and the reason is that using Kerr solution as
seed, the 5-D generated metric inherits second degree polynomials in $x$
and $y$.

\begin{figure}
\centering
\includegraphics[width=8.6cm,height=4cm]{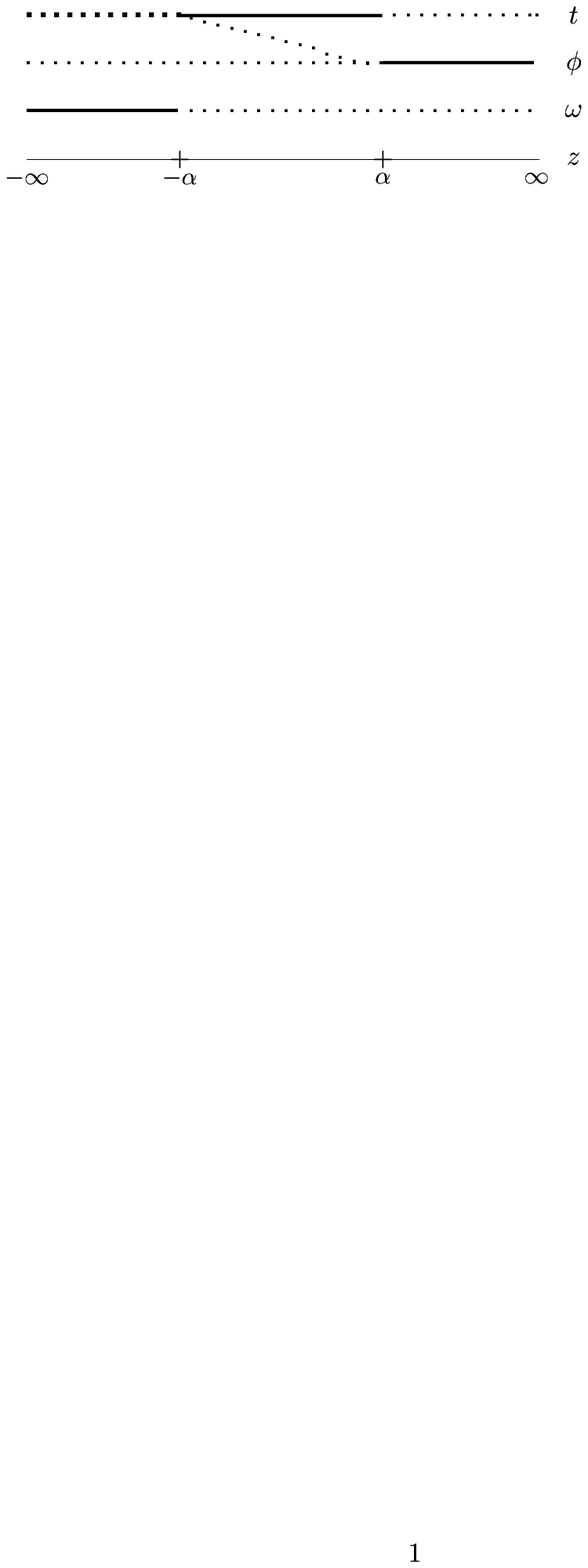}
\caption{\label{fig12}
Rod structure of the 5-D stationary solution generated from Kerr
spacetime by inserting in the direction $\partial_{\omega}$ a
semi-infinite rod $(- \infty, - \alpha)$ with $\Phi= \frac{\sqrt{3}}{2}
\log [\alpha (x+1)(1+y)]$. The bold dotted rod has negative density.}
\end{figure}


\subsubsection{The structure of horizons}

For the stationary solutions, it is interesting to compare the
corresponding scalar $\kappa_{\{t,r\}}$ for the Kerr solution and the
generated solutions. If we choose the fixed coordinates as $\{x^a \}=t, r$
and the coordinates describing the hypersurface as $\{x^A \}=
\theta,\phi$, the scalar $\kappa$ defining the trapping for such spacelike
surface in the Kerr spacetime is given by

\begin{eqnarray}
\kappa^{Kerr}_{\{t,r\}} &&= - g^{rr}(U^{Kerr}_{r})^2 \nonumber\\
&& = - \frac{\Delta}{\Sigma} \frac{[ r(2r^2+a^2+a^2
\cos^2{\theta})+a^2 m \sin^2{\theta}]^2}{[(r^2+a^2)^2- \Delta a^2
\sin^2{\theta}]^2}.
\end{eqnarray}
where $\Delta= r^2- 2mr +a^2$ and $\Sigma= r^2+a^2
\cos^2{\theta}$.

The terms in square brackets are always strictly positive, therefore
$\Delta=0$ determines the only marginally trapped surfaces or horizons
described by the spheres $r_{\pm}= m \pm \sqrt{m^2- a^2}$ corresponding to
the inner and outer horizons in Kerr geometry. When lifted to five
dimensions, the new solution (\ref{Kerr1-5d}) presents, besides the Kerr
horizons, additional marginally trapped surfaces, as can be seen analyzing
the invariant $\kappa$ given by

\begin{eqnarray}
\kappa^{Kerr5}_{\{t,r\}} &&= - \frac{\Delta}{\Sigma}
\frac{(r-m + \alpha \cos{\theta})}{(r-m + \alpha)^{4}
(1+ \cos{\theta})^{2}}
\frac{[{\cal U}(r, \theta)]^2}{[(r^2+a^2)^2 - \Delta
a^2 \sin^2{\theta}]^2}, \nonumber\\
{\cal U}(r, \theta) &&= \frac{3 \alpha}{2}
(\cos{\theta}-1)[(r^2+a^2)^2 - \Delta
a^2 \sin^2{\theta}] +\nonumber\\
&&( r-m + \alpha)(r-m + \alpha \cos{\theta})
[r(2r^2+a^2+a^2\cos^2{\theta})+a^2 m \sin^2{\theta}].
\label{k_kerr5d}
\end{eqnarray}

\begin{figure}
\centering
\includegraphics[width=7cm,height=6.5cm]{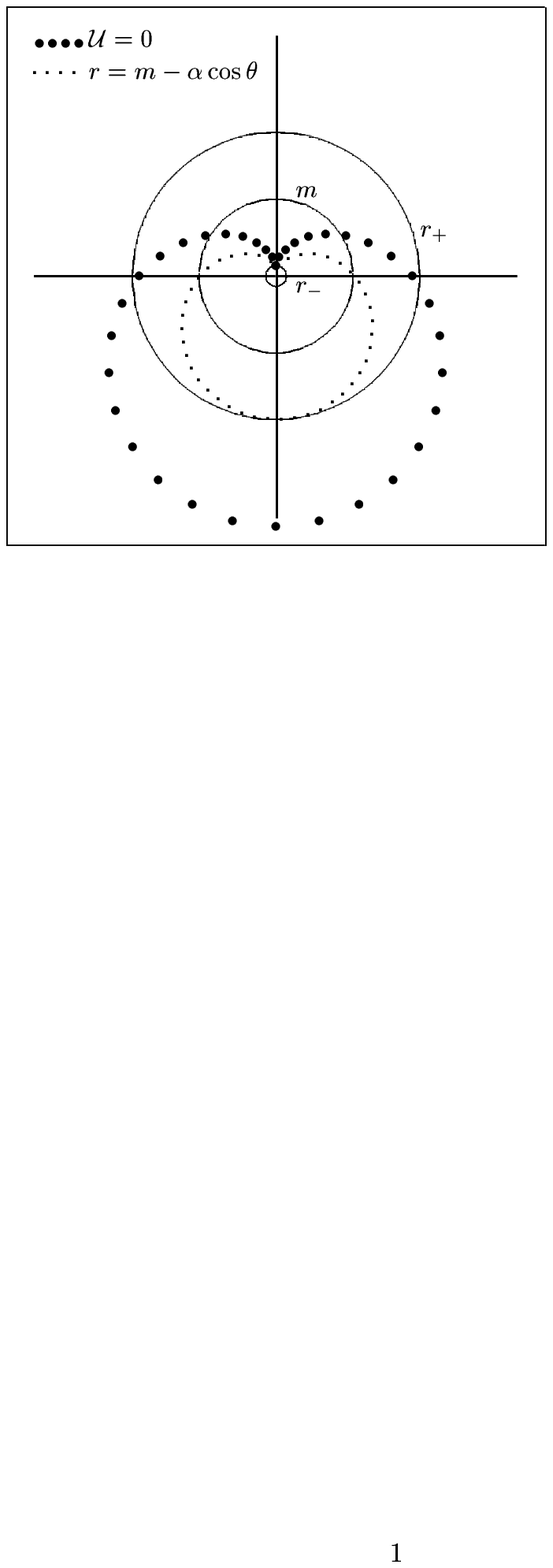}
\caption{\label{fig13}
Horizon squeme for the 5-D stationary solution
generated from Kerr spacetime adding a semi-infinite rod
$(-\infty, - \alpha)$, with
$\Phi= \frac{\sqrt{3}}{2} \log[\alpha (x+1)(1+y)]$.}
\end{figure}

The last expression (\ref{k_kerr5d}) shows that $\kappa_{\{t,r\}}$
vanishes when $\Delta=0$, coinciding with the Kerr inner and outer
horizons; but it also vanishes when $r=r_h=m - \alpha \cos{\theta}$. This
additional marginally trapped surface lies between the inner and outer
Kerr horizons, touching them tangentially. For $r_{h}< m-\sqrt{m^2-a^2}
\cos{\theta}$, $\kappa_{\{t,r\}}$ becomes positive and the surface becomes
trapped.  Furthermore, another marginally trapped surface exists for those
values of $r$ such that ${\cal U}(r, \theta)=0$. Since ${\cal U}$ is a
fifth degree polynomial in $r$ it must have at least one real root
generating a marginally trapped surface. When $a=0$, the surface ${\cal
U}=0$ corresponds to the one found in the static example of the previous
section at $r = \frac{19}{16}m(1- \cos{\theta})$. In Fig. \ref{fig13} a
numerical profile of ${\cal U}=0$ is shown along with slices of other
horizons.


\begin{figure}[ht]
\centering
\includegraphics{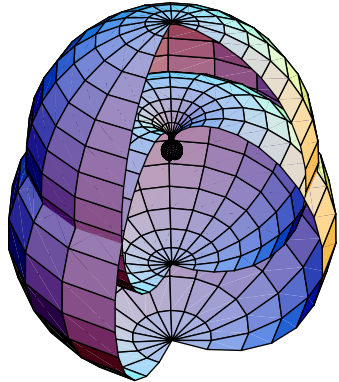}
\caption{Marginally trapped surfaces of Kerr1 5-D, generated with a Kerr
4-D seed plus a dilaton field $\Phi= \frac{\sqrt{3}}{2}\log[(x+1)(1+y)]$,
these are the rotated slices of Fig. \ref{fig13}.}
\label{fig14}
\end{figure}

We now compare these results with those obtained for $\kappa_{\{t,r\}}$
corresponding to the Myers-Perry five-dimensional spinning black hole
\cite{Myers}. The Myers-Perry metric with one rotation $a_1 \ne 0$, in
Boyer-Lindquist coordinates is

\begin{eqnarray}
ds^2&&= -dt^2+ \frac{r_0^2}{\Sigma}[dt-a_1 \sin^2{\theta} d \phi]^2
+(r^2+a_1^2) \sin^2{\theta} d \phi^2 \nonumber\\
&& + r^2 \cos^2{\theta} d \psi^2 + \Sigma \left( \frac{d
r^2}{\Delta} + d \theta^2 \right),
\end{eqnarray}
where $\Sigma= r^2+a_1^2 \cos^2 {\theta}$ and $ \Delta= r^2+a_1^2-
r_0^2$. While the scalar $\kappa_{\{t,r\}}$ is given by

\begin{equation}
\kappa^{MP}_{\{t,r\}}
= - \frac{\Delta}{\Sigma} \frac{[r^2(r^2+a_1^2)
+(2r^2+a_1^2)(r^2+a_1^2 \cos^2{\theta})+r_0^2a_1^2
\sin^2{\theta}]^2}{r^2[(r^2+a_1^2
\cos^2{\theta})(r^2+a_1^2)+r_0^2 a_1^2
\sin^2{\theta}]^2}.
\label{k_kerr5d2}
\end{equation}

The expression (\ref{k_kerr5d2}) clearly diverges at $r=0$ which
corresponds to a strong curvature singularity, while the unique horizon is
given by $\Delta=r^2+a_1^2-r_0^2=0$, and is described by the sphere $r=
\sqrt{r_0^2-a_1^2}$.

\subsection{Second stationary example}

\begin{figure}
\centering
\includegraphics[width=8.6cm,height=4cm]{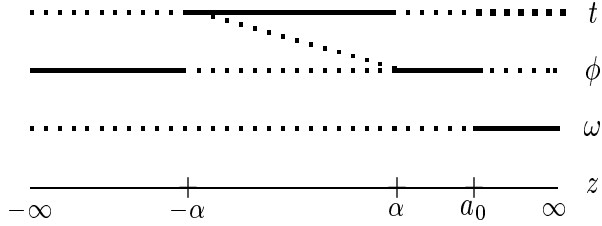}
\caption{\label{fig15}
Rod structure of the 5-D stationary solution generated using Kerr
spacetime as seed and  inserting in the direction $\partial_{\omega}$
a semi-infinite rod $(a_0, \infty)$.}
\end{figure}

Another interesting example is obtained by taking the Kerr solution and
introducing a semi-infinite rod $(a_0, \infty)$ with $\Phi=
\frac{\sqrt{3}}{2} \log[a_0 (x+1)(1-y)]$. It follows that the rod
structure resembles the Emparan-Reall black ring \cite{emparan}, as shown
in Fig. \ref{fig15}. The metric is given by the following expression

\begin{eqnarray}
ds^2&&= e^{\sigma}\left( \frac{dx^2}{x^2-1}+ \frac{dy^2}{1-y^2}  \right) +
\gamma_{ab}dx^a dx^b, \quad a,b=t, \phi, \omega,
\nonumber\\
e^{\sigma}&&= \frac{m^2}{a_0} \frac{(x+1)^2(1-y)^2}{(x-y)^3}
[(1+px)^2+q^2y^2], \nonumber\\
g_{tt}&&=- \frac{1}{a_0(x+1)(1-y)}
\frac{(p^2x^2+q^2y^2-1)}{[(1+px)^2+q^2y^2]}, \nonumber\\
g_{t \phi}&&=- \frac{2a(1+y)}{a_0(x+1)}
\frac{(1+xp)}{[(1+px)^2+q^2y^2]}, \nonumber\\
g_{\phi \phi}&&=- \frac{(1+y)}{a_0(x+1)}
\frac{ \{ 4a^2(1-y^2)(1+xp)^2-
\alpha^2(x^2-1)[(1+px)^2+q^2y^2]^2
\}}{[(1+px)^2+q^2y^2][p^2x^2+q^2y^2-1]},
\nonumber\\
g_{\omega  \omega}&&=[a_0(x+1)(1-y)]^2.
\label{E-R}
\end{eqnarray}

Finally the scalar $\kappa_{\{t,r\}}$ for the solution (\ref{E-R})
amounts to

\begin{eqnarray}
\kappa_{\{t,r\}} &
=& -\frac{a_0m^2}{ \alpha} \frac{\Delta}{\Sigma} \frac{(
r-m- \alpha \cos{\theta})}{(r-m+ \alpha )^4(
1- \cos {\theta})^2}
\frac{G(r, \theta)^2}{[(r^2+a^2)^2- \Delta a^2 \sin^2{\theta}]^2},
\nonumber\\
G(r, \theta)&=&
\frac{3 \alpha}{2} (\cos{\theta}+1)[(r^2+a^2)^2 - \Delta
a^2 \sin^2{\theta}] - \nonumber\\
&&(r-m + \alpha)(r-m - \alpha \cos{\theta})
\times \nonumber\\
&&[r(2r^2+a^2+a^2\cos^2{\theta})+a^2 m \sin^2{\theta}].
\end{eqnarray}

\begin{figure}
\centering
\includegraphics[width=7cm,height=6.5cm]{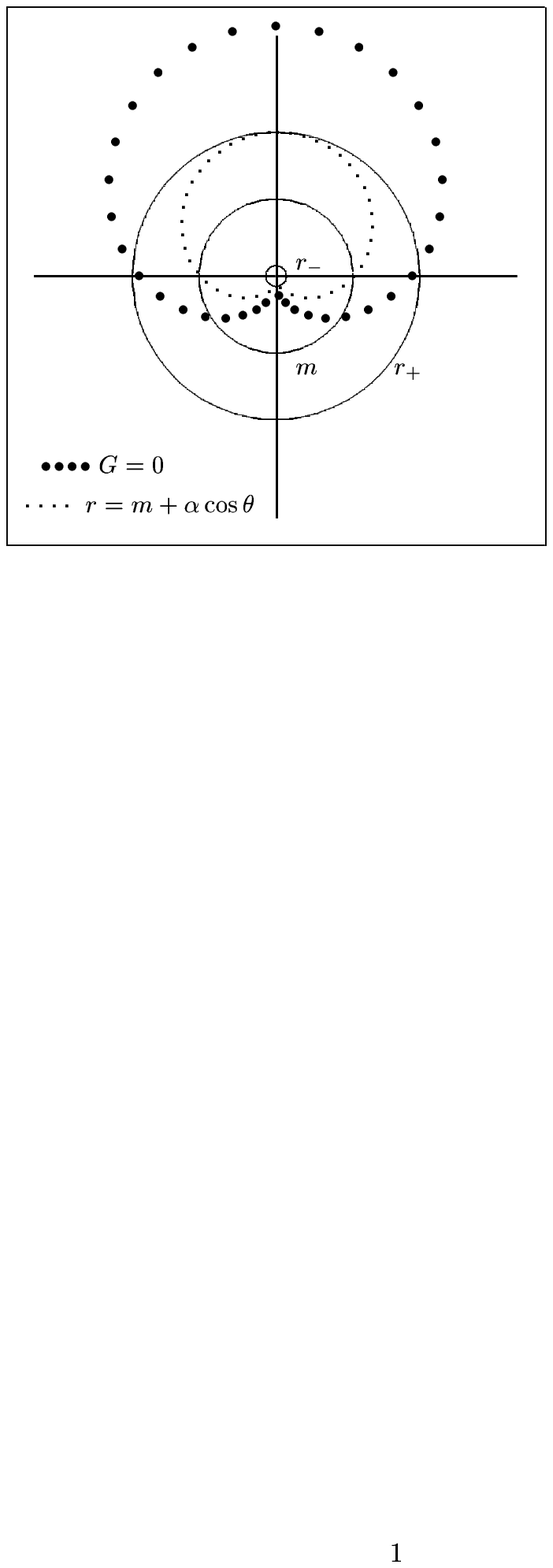}
\caption{\label{fig16}
Horizon slices for the stationary solution
generated from Kerr spacetime adding a semi-infinite rod $(a_0, \infty)$;
$r_{\pm}=m \pm \sqrt{m^2-a^2}$.}
\end{figure}

The corresponding rod structure resembles the one of a black ring due to
Emparan-Reall \cite{emparan}, however, the horizons rather correspond to
the 5-D rotating black hole and not the ring. The marginally trapped
surfaces are located at $\Delta=0$ and $(r_{h}-m- \alpha \cos{\theta})=0$,
or $r_{\pm}=m \pm \sqrt{m^2-a^2}$ and $r_{h}= m+ \sqrt{m^2-a^2}
\cos{\theta}$. A new marginally trapped surface arises from $G(r,
\theta)=0$ and has the shape of the previously studied ${\cal U}=0$ (Fig.
\ref{fig13}), but turned upside down. The profiles of horizons are shown
in Fig. \ref{fig16}.

The generated stationary metrics (\ref{Kerr1-5d}) and (\ref{E-R}) acquire
a simpler form in Boyer-Lindquist coordinates, the expressions are
presented in the Appendix C.


\section{Conclusions}

In this paper we addressed the construction in a simple and controllable
way of higher dimensional vacuum solutions to Einstein equations which
depend, at most, on two variables. The four dimensional vacuum seed
metrics with $G_{2}$ symmetry are generalized to include massless
dilatons, which serve to lift the solutions to higher dimensions.  The
method works both, in the case when the two commuting Killing vectors of
the ``seed" are spacelike as well as when one of the Killing vectors is
timelike. The ``translation" of the algorithm to the case of one spacelike
and one timelike Killing vector is given here for the first time.

The algorithm is illustrated with the generation of some static and
stationary solutions.  Starting with Minkowski, Schwarzschild and Kerr
seeds and ``adding" rods to the fifth dimension, deformed 5-D black hole
solutions were generated. The corresponding rod structure of some of these
solutions resembles the Myers-Perry black hole or the Emparan-Reall black
ring, however the topology of the horizons is rather different. The
generated solutions present distorted horizons due to the presence of
extra sources.
  
We have seen that the rod structure alone does not reflect some important
properties of spacetime, since it is insensitive to the exponents or
powers of the metric functions. Neither, rod directions are apparent from
the metric expressions in the sense that even for static metrics the rods
may have crossed components aligned with spacelike and timelike Killing
directions. Therefore in order to characterize these spacetimes one must
perform the singularity analysis and study their horizons. Nevertheless,
by imposing conditions on asymptotic behavior of spacetimes as well as
certain physical properties one may show \cite{hollands} how the rod
structure is important to single out such a spacetime.

In future works it would be interesting to consider five dimensional
solutions with double rotation. Such solutions can be obtained by lifting
stationary solutions with both electromagnetic and scalar fields. In this
case the second rotation will be induced by the crossed terms which appear
due to the lifting of the electromagnetic degrees of freedom. These
solutions will be discussed elsewhere.

\appendix
\section{Einstein-scalar
equations in Weyl coordinates}

Here we present for completeness the Einstein-scalar coupled field
equations in Weyl coordinates. The general stationary axisymmetric line
element can be put:

\be
ds^2= -e^{U} (dt+ {\rm A}d\phi)^2+e^{-U}r^2d\phi^2+e^{\sigma}(dr^2+dz^2).
\label{axi}
\ee

The  Einstein field equations are:

a) The $U$-{\rm A} equations
\be
U_{rr} + (1/r)\,U_{r}+U_{zz}+\frac{e^{2U}}{4r^{2}}\,\left[{\rm
A}_{r}^2+{\rm A}_{z}^2\right]=0, \label{U-A1}
\ee

\be
\left(\frac{e^{2U}{\rm A}_{r}}{r}\right)_{r}+
\left(\frac{e^{2U}{\rm A}_{z}}{r}\right)_{z}=0, \label{U-A2}
\ee

b) The $\Phi$ equation
\be
\Phi_{rr} + \frac{1}{r}\,\Phi_{r}+\Phi_{zz}=0, \label{KGAp}
\ee
and finally

c) The $\sigma$ equation
\be
\sigma_{r}+U_{r}= r\left[\Phi_{r}^{2}-\Phi_{z}^{2}\right]
+\frac{r}{2}\left[U_{r}^{2}-U_{z}^{2}\right]-\frac{e^{2U}}{2r}
\left[{\rm A}_{r}^{2}-{\rm A}_{z}^{2}\right],
\label{frA}
\ee

\be
\sigma_{z}+ U_{z} =2\,r\Phi_{r}\Phi_{z}+
\,rU_{r}U_{z}-\frac{e^{2U}}{r}{\rm A}_{r}{\rm A}_{z}. \label{fzA}
\ee

When ${\rm A}=0$ we deal with static solutions. Notice that the
contribution of the function $U$ to the non-linear $\sigma$ is identical
to that of $\Phi$. The solutions are defined by $U$, $\Phi$ and A, while
the function $\sigma$ is obtained by quadratures.

\section{Analysis of the rod structure for Kerr 5-D solution
(\ref{Kerr1-5d}).}

To determine the direction of each rod we follow the steps of Sec.III in
\cite{harmark}, for the solution (\ref{Kerr1-5d}). The equation to analyze
for each interval on the z-axis is $g_{ij} \vec{v}=0$, explicitly:

\begin{equation}
e^{- \frac{2 \Phi}{ \sqrt{3}}}
\left( \begin{array}{ccc}
{\gamma}_{tt}&
{\gamma}_{t \phi}& 0\\
{\gamma}_{\phi t}&
{\gamma}_{\phi \phi}& 0\\
0&0&e^{\frac{6 \Phi}{ \sqrt{3}}}
\end{array} \right)
\left( \begin{array}{c}
v^1 \\ v^2 \\ v^3
\end{array} \right)=0,
\end{equation}
where $\Phi=\frac{\sqrt{3}}{2} \log [\alpha (x+1)(1+y)]$ is the introduced
rod, while $\vec{v}$ is the direction of the rod corresponding to the
analyzed interval and ${\gamma}_{ab}$ denote the seed metric functions.
The analysis is performed in the limit $r \to 0$ that is

\begin{equation}
x= \frac{ |z+ \alpha|+|z- \alpha|}{2 \alpha}, \quad
y= \frac{| z+ \alpha|- |z- \alpha|}{2 \alpha}.
\end{equation}

The rod structure is as follows:
(i) The semi-infinite spacelike rod $z \in (- \infty, - \alpha)$
corresponds to
$x= - \frac{z}{\alpha}$ and $y=-1$. Substituting in
(\ref{Kerr1-5d}), we get

\begin{eqnarray}
g_{tt}&&= \frac{p^2 (1+ \frac{z}{\alpha})}{\alpha (1+y)
[(1-\frac{pz}{\alpha})^2+q^2]}, \nonumber\\
g_{t \phi}&&=- \frac{4a}{ \alpha}
\frac{(1-\frac{pz}{\alpha})}
{(1-\frac{z}{\alpha})[(1-\frac{pz}{\alpha})^2+q^2]},
\nonumber\\
g_{\phi \phi}&&= \frac{2 \alpha}{p^2}
\frac{[(1- \frac{pz}{\alpha})^2+q^2]}{(1-\frac{z}{\alpha})},
\nonumber\\
g_{\omega  \omega}&&=\alpha^2 (1-\frac{z}{\alpha})^2(1+y)^2=0.
\label{Kerr1-5d-r=0}
\end{eqnarray}
Note that $g_{\omega \omega}=0$ and $g_{tt}$ diverges in this interval.
The vanishing of $g_{\omega \omega}$ means that the rod $(- \infty, -
\alpha)$ is entirely located along the $\partial_{\omega}$ direction, i.e.
$v^1=v^2=0, \quad v^3=1$.

The analysis is analogous for the rod $z \in (\alpha, \infty)$ that
corresponds to $x= \frac{z}{\alpha}$ and $y=1$. Substituting in
(\ref{Kerr1-5d}) we obtain that $g_{tt} \ne 0$, $g_{t \phi}=g_{\phi
\phi}=0$ while $g_{\omega \omega} \ne 0$. Solving the system $g_{ij}
\vec{v}=0$ gives that the rod $(\alpha, \infty)$ has the direction
$v^{1}=v^3=0$ and $v^2=1$, i. e. it is situated along $\partial_{\phi}$.

(ii) The finite timelike rod $(- \alpha, \alpha)$ corresponds
to $y= \frac{z}{\alpha}$ and $x=1$.
Substituting in (\ref{Kerr1-5d}) we get
\begin{eqnarray}
g_{tt}&&= \frac{q^2 (1- \frac{z}{\alpha})}
{2 \alpha
[(1+p)^2+q^2 \left( \frac{z}{\alpha}\right)^2]}, \nonumber\\
g_{t \phi}&&=- \frac{a}{ \alpha}
\frac{(1+p)(1-\frac{z}{\alpha})}{[(1+p)^2+q^2 \left(
\frac{z}{\alpha}\right)^2]},
\nonumber\\
g_{\phi \phi}&&= \frac{2 a^2 (1+p)^2}{ \alpha q^2}
\frac{(1-\frac{z}{\alpha})}{[(1+p)^2+q^2 \left(
\frac{z}{\alpha}\right)^2]},
\nonumber\\
g_{\omega  \omega}&&= 4 \alpha^2 (1+ \frac{z}{\alpha})^2.
\label{Kerr-5d-timerod}
\end{eqnarray}

Solving the system $g_{ij} \vec{v}=0$ gives that the rod $(\alpha,
\infty)$ has components along the two Killing directions $\partial_{t}$
and $\partial_{\phi}$, $v^{3}=0$, $v^{1}=1$ and $v^2=
\Omega=\frac{q}{2m(1+p)}$. The corresponding rod structure is shown in
Fig. \ref{fig12}.

\section{5-D stationary metrics in Bo\-yer-Lind\-quist co\-ordinates}

The generated stationary metrics with Kerr as seed, expressed in
Boyer-Lindquist coordinates are given below. At the end we include the
Schwarzschild 4-D solution in Weyl coordinates. The transformation between
prolate spheroidal coordinates and Boyer-Lindquist coordinates is
$x=(r-m)/\alpha, \quad y= \cos{\theta}$.

\begin{eqnarray}
ds^2&=& e^{\sigma} \left( \frac{d r^2}{\Delta}+{d \theta^2}
\right) + g_{ij}dx^i dx^j, \quad i,j=t, \phi, \omega,
\nonumber\\
e^{\sigma}&=& \frac{\alpha}{a_0} \Sigma \frac{(r-m+ \alpha)^2(1 \pm
\cos{\theta})^2}{(r -m \pm \alpha \cos{\theta})^3},
\nonumber\\
g_{tt}&=&- \frac{\alpha}{a_0}
\frac{( \Delta- a^2 \sin^2{\theta})}{\Sigma (r -m+\alpha)(1 \pm
\cos{\theta})},
\nonumber\\
g_{t \phi}&=&- \frac{2am \alpha(1 \mp \cos{\theta}) r}{a_0(
r -m+\alpha) \Sigma}, \nonumber\\
g_{\phi \phi}&=&- \frac{\alpha(1 \mp \cos{\theta})}{a_0(r -m+\alpha)
\Sigma} [(r^2 +a^2)^2- \Delta a^2 \sin^2{\theta}],
\nonumber\\
g_{\omega  \omega}&=&
\frac{a_0^2}{\alpha^2}(r -m+\alpha)^2(1 \pm \cos{\theta})^2,\nonumber\\
\Delta &=& r^2 -2mr +a^2, \quad \Sigma= r^2+a^2 \cos^2{\theta}.
\label{Kerr-5d-BL}
\end{eqnarray}

The upper sign corresponds to a rod structure of Kerr when the
semi-infinite rod $(- \infty, - a_0)$ is inserted, while the lower sign is
for the spacetime generated from Kerr with the inserted semi-infinite rod
$(a_0, \infty)$, both with the scalar charge $A= \frac{ \sqrt{3}}{2}$. In
the first case we took $a_0 = \alpha$.

The Schwarzschild solution in Weyl coordinates is given by the line
element:

\begin{eqnarray}
ds^2&&= -e^{U}dt^2+ e^{\sigma^{\rm vac}}(dr^2+dz^2)+ r^2 e^{-U}d
\phi^2,
\nonumber\\
\sigma^{\rm vac} &&=-U+\gamma, \nonumber\\
U&&=-  \log \left[ \frac{m-z+ r_{-}}{-m-z+
r_{+}} \right],\nonumber\\
\gamma &&= \log \left[ \frac{(r_{-} +
r_{+})^2-4m^2}{4r_{-}r_{+}}
\right],
\label{Schw-4d,Weyl}
\end{eqnarray}
where $r_{\pm}^2=(m \pm z)^2+r^2$.

\begin{acknowledgments}
A.F.  aknowledges the support of the Basque Government Grant
GICO7/51-IT-221-07 and The Spanish Science Ministry Grant FIS2007-61800.
N. B. would like to thank the colleagues of UPV/EHU for warm hospitality.
L. A. L\'opez acknowledges Conacyt-M\'exico for a Ph. D. grant. Partial
support of Conacyt-Mexico Project 49182-F is also acknowledged.
\end{acknowledgments}

\end{document}